\documentclass[12pt]{article}

\usepackage{amsmath,amssymb,amsfonts,amsbsy}
\usepackage{cite}
\usepackage{graphicx}
\usepackage{wrapfig}
\usepackage{epsfig}
\begin{document}

%%%%%%% please do not touch these! %%%%%%
\setcounter{section}{0} \setcounter{subsection}{0} \setcounter{equation}{0}
\setcounter{figure}{0} \setcounter{footnote}{0} \setcounter{table}{0}

\begin{center}
\textbf{THE FIRST STAGE OF  POLARIZATION PROGRAM SPASCHARM AT THE ACCELERATOR
U-70 OF IHEP}

\vspace{5mm} V.V.~Abramov$^{\,1}$, N.A.~Bazhanov$^{\,2}$, 
N.I.~Belikov$^{\,1}$, A.A.~Belyaev$^{\,3}$, A.A.~Borisov $^{\,1}$, 
N.S.~Borisov$^{\,2}$, M.A.~Chetvertkov$^{\,4}$, V.A.~Chetvertkova$^{\,5}$, 
Yu.M.~Goncharenko$^{\,1}$, V.N.~Grishin $^{\,1}$, A.M.~Davidenko $^{\,1}$,
A.A.~Derevshchikov $^{\,1}$, R.M.~Fahrutdinov$^{\,1}$,V.A.~Kachanov $^{\,1}$,
Yu.D.~Karpekov$^{\,1}$, Yu.V.~Kharlov$^{\,1}$, V.G.~Kolomiets$^{\,2}$, 
D.A.~Konstantinov$^{\,1}$, V.A.~Kormilitsyn$^{\,1}$, A.B.~Lazarev$^{\,2}$, 
A.A.~Lukhanin$^{\,3}$, Yu.A.~Matulenko$^{\,1}$, Yu.M.~Melnik$^{\,1}$, 
A.P.~Meshchanin$^{\,1}$, N.G.~Minaev$^{\,1}$, V.V.~Mochalov$^{\,1}$, 
D.A.~Morozov$^{\,1}$, A.B.~Neganov$^{\,2}$, L.V.~Nogach$^{\,1}$, 
\underline{S.B.~Nurushev}$^{\,1\,\dag}$, V.S.~Petrov$^{\,1}$, 
Yu.A.~Plis$^{\,2}$, A.F.~Prudkoglyad$^{\,1}$, A.V.~Ryazantsev$^{\,1}$, 
P.A.~Semenov$^{\,1}$, V.A.~Senko$^{\,1}$, N.A.~Shalanda$^{\,1}$, 
O.N.~Shchevelev$^{\,2}$, L.F.~Soloviev$^{\,1}$, Yu.A.~Usov$^{\,2}$, 
A.V.~Uzunian$^{\,1}$, A.N.~Vasiliev$^{\,1}$, V.I.~Yakimchuk$^{\,1}$, 
A.E.~Yakutin$^{\,1}$

\vspace{5mm}

\begin{small}
(1) \emph{IHEP, Protvino, Russia} \\
  (2) \emph{JINR, Dubna, Russia} \\
  (3) \emph{KhPTI, Kharkov, Ukraine}\\
  (4) \emph{MSU, Physics Department, Moscow, Russia} \\
  (5) \emph{Skobeltsyn INP MSU, Moscow, Russia} \\

$\dag$ \emph{E-mail: Sandibek.Nurushev@ihep.ru}

\end{small}
\end{center}

\vspace{0.0mm} % Don't laugh: it does change the spacing!

\begin{abstract}
  \par
The first stage of the proposed polarization program SPASCHARM includes the measurements of the single-spin asymmetry (SSA) in exclusive and inclusive reactions with production of stable hadrons and the light meson and baryon resonances.In this study we foresee of using the variety of the unpolarized beams ( pions, kaons, protons and antiprotons) in the energy range of 30-60 GeV.  The polarized proton and deuteron targets will  be used for revealing the flavor and isotopic spin dependencies  of the polarization phenomena. The neutral and charged particles in  the final state will be detected.
\end{abstract}
\vspace{7.2mm}
\begin{center}
\section*{Introduction}
\end{center}
In shaping the new polarization program at U-70 we
were guided by three conditions:  by our own
experiences. by theoretical status of subject and the reliability of the new program.As concerns of the first condition one may refer  
 on the comparative study of
polarizations in the elastic scattering of particles and antiparticles by
using the polarized proton target \cite{gaidot1}, \cite{gaidot2},
measurements of the spin transfer tensor \cite{pier1}, \cite{pier2} (HERA
Collaboration), study of the SSA in the exclusive and inclusive charge exchange reactions at 40 GeV/c \cite {mochal} (PROZA Collaboration), study of polarization effects at 200 GeV/c by using the polarized proton and antiproton beams (E581/E704 Collaboration, FNAL) \cite {nurush}. The fourth example of  polarization data came recently from the STAR Collaboration at energy in the center of mass $\sqrt{s}$=200 GeV \cite{nogach}, \cite{xu}.\\
The second condition is the  status of the relevant theoretical models. Since
the  energies and transfer momenta with which we are dealing are not large
enough, so there is a doubt about the possible application of the perturbative
quantum chromodynamics  p(QCD). Therefore either the specific models should
be used for the interpretation of the experimental data or the general
asymptotic predictions might be applied.
\par
The third condition is relevant to the reliability of the experiment, that
is, availability of robust equipments, manpower, money and other resources.
\par
Below we shall briefly describe all these conditions.
%\begin{center}
\section{The preceding study of polarization effects at U70}
%\end{center}
\par
\begin{figure}[b!]
  \centering
  \begin{tabular}{ccc}
  %\textmd{\textit{\textmd{\begin}}}{tabular}{ccc}
    \includegraphics[width=50mm]{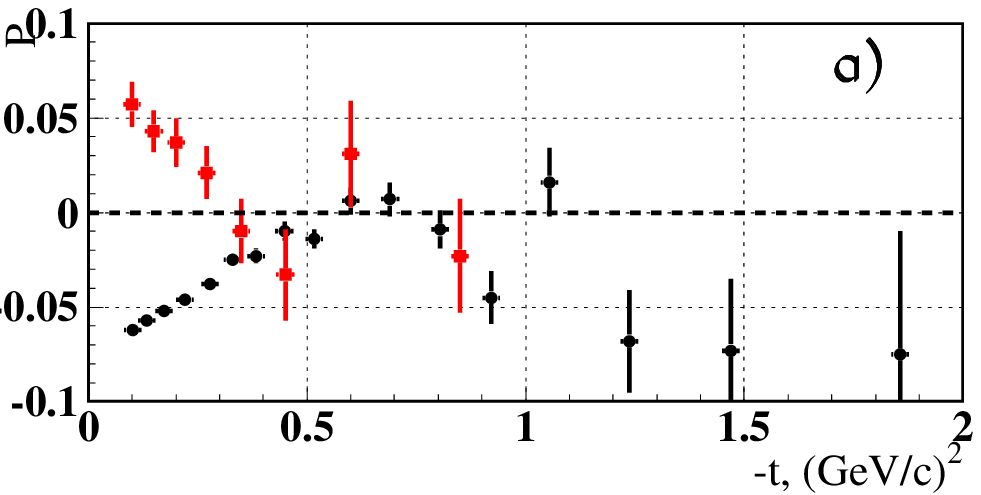} &
    \includegraphics[width=50mm]{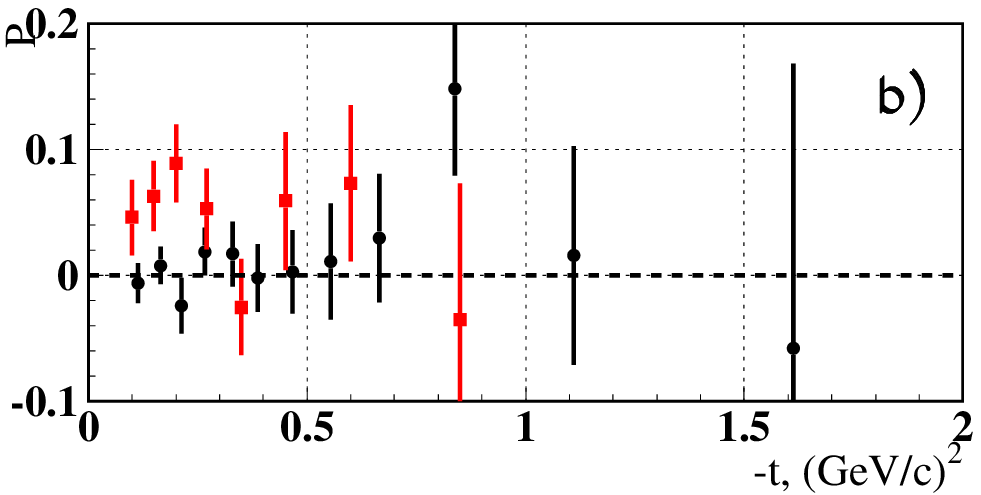} &
    \includegraphics[width=50mm]{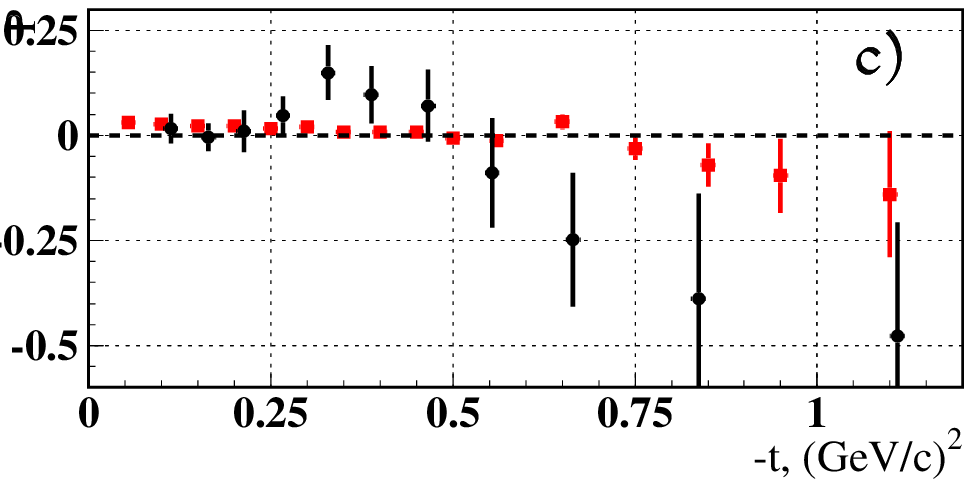} \\
    \textbf{(a)} &   \textbf{(b)} &  \textbf{(c)}
  \end{tabular}
  \caption{%
    \textbf{(a)}P in elastic scattering:$\pi^{-}p(\bullet)$ and $\pi^{+}p(\blacksquare)$.
    \textbf{(b)}P in elastic scattering:$K^{-}p(\bullet)$ and $K^{+}p(\blacksquare)$ .
    \textbf{(c)}P in elastic scattering:$\bar{p}p(\bullet)$ and $pp(\blacksquare)$.
    }
  \label{Fig.1}
\end{figure}

In 1970-1976 at U70 Collaboration of physicists from Saclay (France), Protvino, Dubna and Moscow (Russia)  (HERA Collaboration) had performed the measurements of the polarization parameters P and R (spin rotation parameter) in elastic scattering of particles and antiparticles at $\sim$ 40 GeV/c by using the polarized proton target.  The polarization data are presented in Figure \ref{Fig.1} with one panel for pair of particle and antiparticle.\\
In  papers \cite{logun} and  \cite{nambu}it was considered some consequences of the hypothesis of the approximate  $\gamma_{5}$ invariance of the strong interactions. According to this hypothesis at high energies and large momentum   transfers s,  -t $\gg m^{2}$  (m is the mass of the particles involved in reactions)    the polarization in any elastic scattering of particles or antiparticles should be equal to zero. From Figure \ref{Fig.1} the following results stem out of:
 \begin{enumerate}
   \item polarizations are not zero in reactions induced by pions, protons and antiprotons. It means that the hypothesis of  $\gamma_{5}$  invariance does not work for that reactions yet,
   \item the polarizations are zero for reactions induced by kaons. It means that the hypothesis of  $\gamma_{5}$  invariance   may work in these reactions. But the large error bars in the measured polarizations make this statement doubtful. The future experiments measuring the polarizations in kaon induced elastic scattering with better statistics are needed.
        \end{enumerate}

 For all of above reactions the next comment follows. Though the asymptotic regime was reached for s it's not fulfilled for t, since for $\mid t \mid >1(GeV/c)^{2}$ the experimental errors are large. This is the next item for the future measurements with high statistics. There is a good measurement of the polarization parameter in $\pi^{\pm}p, K^{\pm}p, pp, \bar {p} p$ elastic scattering at 6 GeV/c \cite{borghini}. In this case the $\gamma_{5}$ invariance does not work too for all reactions, exception is $\bar {p}p$, where polarization is compatible with zero in small -t region in frame of the large error bars.\\
In paper \cite{bil} the study was made of the  asymptotic relations between
polarizations in cross channels of a reaction. Using the crossing symmetry
and Fragman - Lindeloff theorem  they   arrived at the following result:
polarizations in the elastic scattering processes (see Figure \ref{Fig.1})  induced by
particle and antiparticle at a given energy and a given angle should be equal
in magnitude and opposite in sign. If we look at Figure \ref{Fig.1} one may note that this
statement is approximately correct for the pion induced reactions, not
correct for reactions initiated by proton and antiproton and
  ambiguous for reactions involving kaons (thanks to the small statistics). Therefore the new elastic scattering experiment should clarify this interesting problem by gathering a large statistics, specially at large transverse momenta.\\
The HERA Collaboration making use of the simple Regge pole model concluded
that the elastic scattering polarizations induced by pions and kaons follow
the predictions of such model, while polarization in elastic pp scattering
reveals the drastic deviation from the prediction of the Regge pole model.
Such behavior may be explained by assuming that at 40 GeV/c momentum the
dominant contribution to the polarization in elastic pp scattering comes from
the pomeron with the spin flip term of
the order of $10\%$ with respect to the spin non flip term. Involving in the analysis the data on the spin rotation parameter \cite{pier1}, \cite{pier2} they strengthened their conclusion. But the statistics are not so large to be unambiguous in such conclusion. One needs more measurements.\\

\begin{figure}[b!]
  \centering
   \begin{tabular}{ccc}
  %\textmd{\textit{\textmd{\begin}}}{tabular}{ccc}
   \includegraphics[width=50mm,height=50mm]{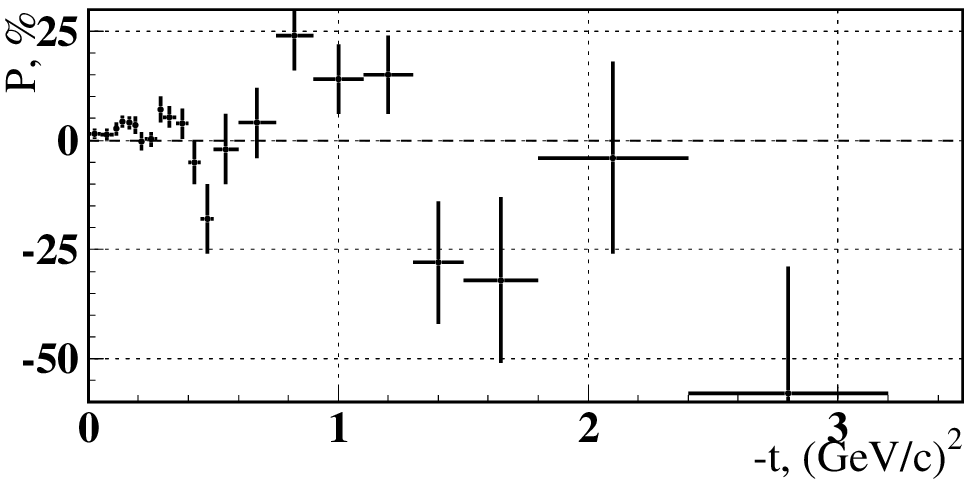} &
   \includegraphics[width=50mm,height=50mm]{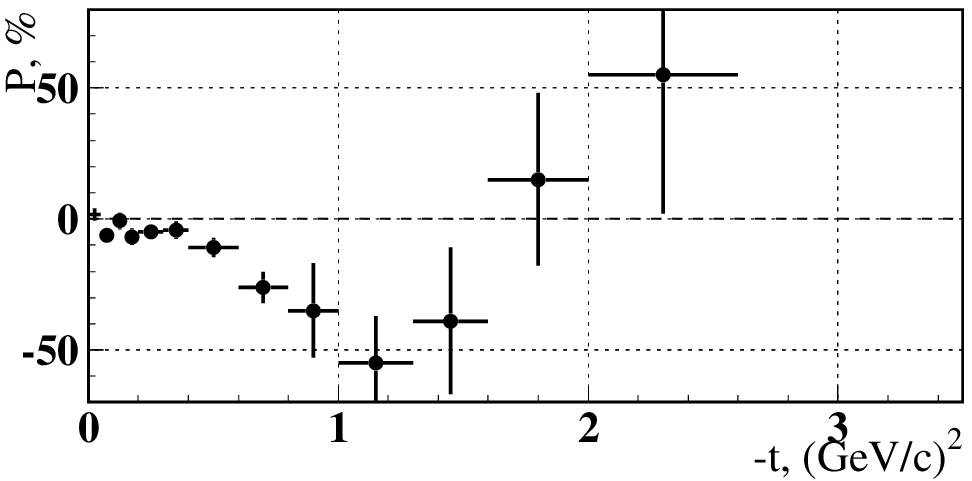} &
   \includegraphics[width=50mm,height=50mm]{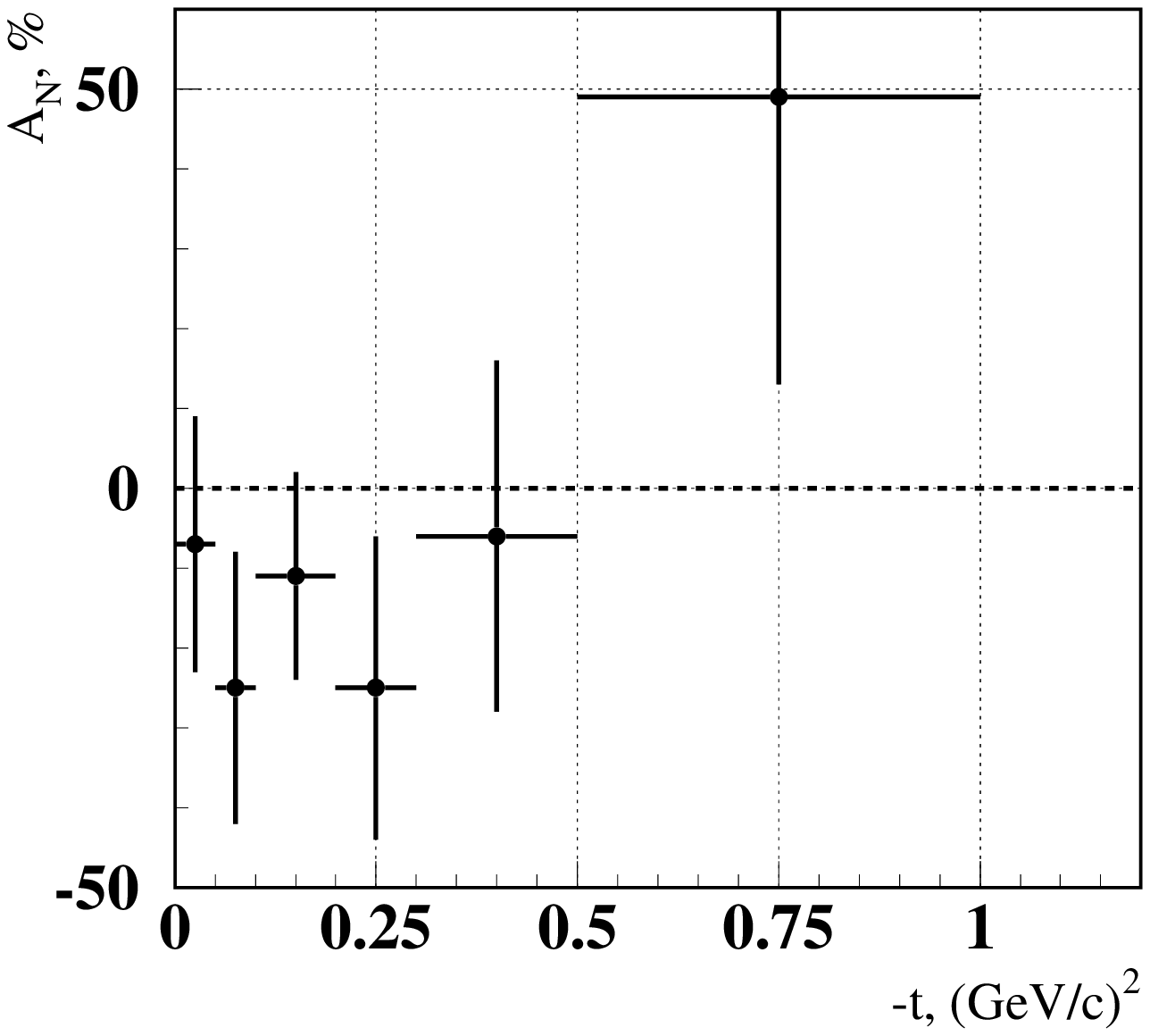} \\
    \textbf{(a)} &
    \textbf{(b)} &
    \textbf{(c)}
  \end{tabular}
  \caption{%
    \textbf{(a)} Polarization in reaction $\pi^{-}+p\rightarrow\pi^{0}+n$.
    \textbf{(b)}Polarization in reaction $\pi^{-}+p\rightarrow\eta+n$ .
    \textbf{(c)}Polarization in reaction  $\pi^{-}+p\rightarrow\eta'+n$.
    }
  \label{Fig.2}
\end{figure}

\begin{figure}[b!]
  \centering
     \begin{tabular}{ccc}
  %\textmd{\textit{\textmd{\begin}}}{tabular}{ccc}
   \includegraphics[width=50mm,height=50mm]{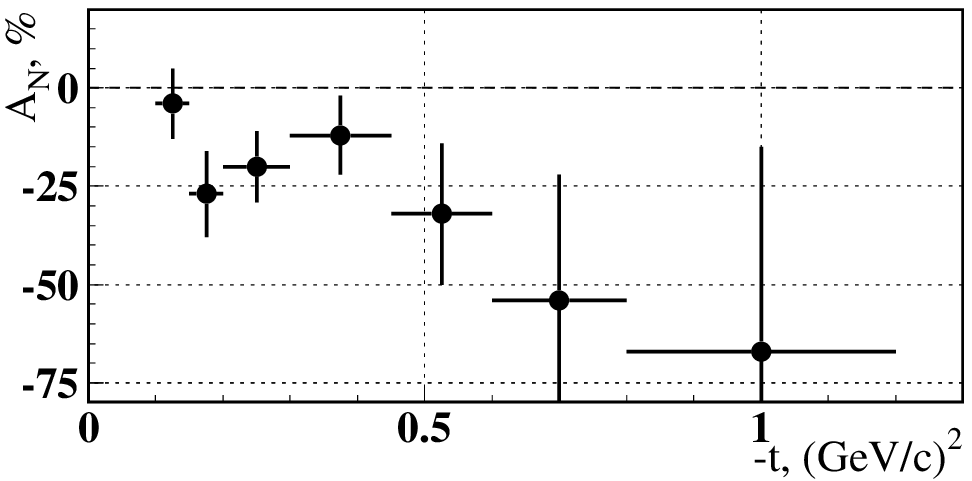} &
   \includegraphics[width=50mm,height=50mm]{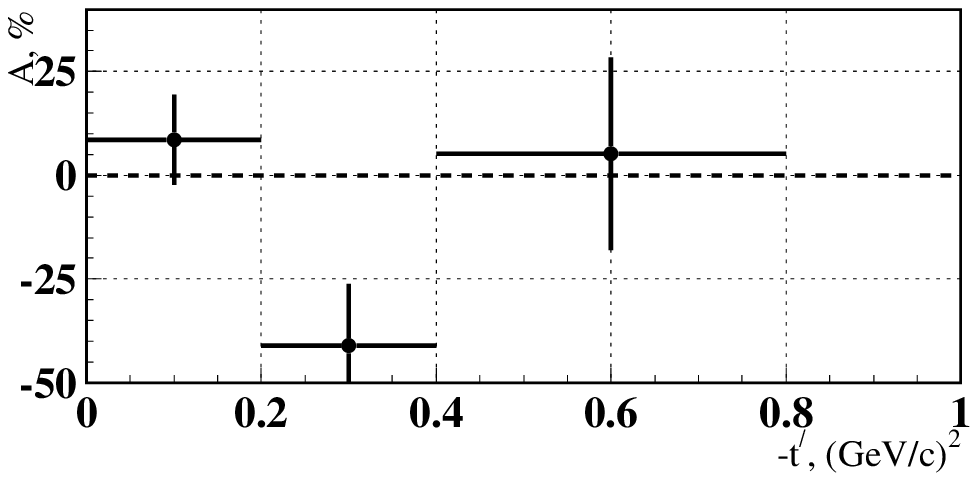} &
    \includegraphics[width=50mm,height=50mm]{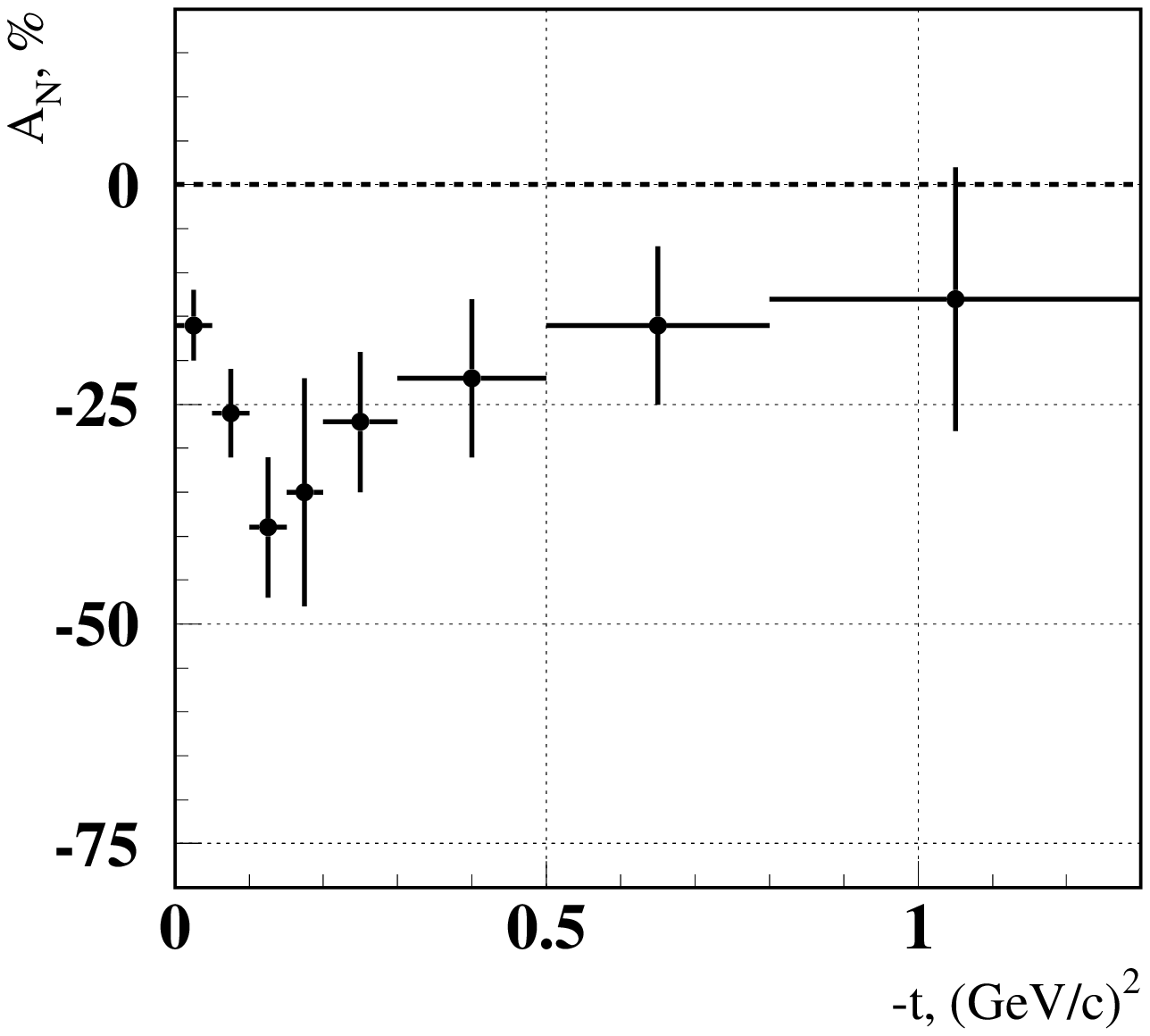} \\
    \textbf{(a)} &
    \textbf{(b)} &
    \textbf{(c)}
  \end{tabular}
  \caption{%
    \textbf{(a)} asymmetry in reaction $\pi^{-}+p\rightarrow\omega+n$.
    \textbf{(b)}asymmetry in reaction$\pi^{-}+p\rightarrow a_2+n$ .
    \textbf{(c)}asymmetry in reaction $\pi^{-}+p\rightarrow f+n$.
    }
  \label{Fig.3}
\end{figure}

The PROZA Collaboration measured the single spin asymmetries in the charge exchange binary and inclusive reactions at the incident beam momentum 40 GeV/c \cite{mochal}. With the different statistics the results were obtained for the  exclusive reactions containing in the final states the mesons of the different mass and quantum numbers Figure \ref{Fig.2} , Figure \ref{Fig.3}.\\

%%%%%%%%%%%%%%%%%%%%%%%%

The polarization data for reactions 1, 2 and 3 (the mesons in the final states
are spinless) were extensively analyzed in frame of the different models. For
example, in the asymptotic model \cite{bil} the polarizations in all of these
three reactions should be zero. But such predictions are in contrast to the
experimental data (see Figure \ref{Fig.2} ). In the Regge pole model with inclusion of the
odderon \cite{enko}, \cite{gauron}, the best approximation predicting the
new dip in polarization around the crossing point at -t$\sim 0.2 (GeV/c)^{2}$
was obtained in the model \cite{enko}. After analyzing the reaction (Figure \ref{Fig.2}a)
the authors of the paper \cite{gauron} noted:
\begin{quote}
The surprising results of the recent 40 GeV/c Serpukhov measurement of the
polarization in $\pi^{-}p\rightarrow\pi^{0}n$ are shown to support the
conjecture that the crossing-odd amplitude may grow asymptotically as fast as
is permitted by general principles.
\end{quote}

This model in contrast to other ones predicts the shift of the left zero
crossing point  farther to left and the increase of polarization with
growth of the incident
 momentum. This is an attractive subject for experimental
 check.\\
The reactions 2,3 were also analyzed in frame of the Regge pole model and data are consistent with model prediction. Other reactions 4-6 showing also the significant spin effects (see Figure \ref{Fig.3}) did not yet attract the attention of theoreticians.\\
These data are unique in the sense that nobody made (30 years later) the
similar measurements at higher energies. This fact may confirm the assumption
that the U-70 accelerator occupies a good niche for such studies of exclusive
reactions dying rapidly with growth of energy.

\par
By  using the experimental set-up PROZA the inclusive asymmetries were measured at 40 GeV/c in the following charge exchange reactions: $$ \pi^{-}+p\rightarrow\pi^{0}+X\ (1).  p+p\rightarrow\pi^{0}+X (2)$$ in the central, polarized target  and unpolarized beam fragmentation regions \cite{mochal}.In central region the asymmetry about 30\% was found at $p_{T}>2$ GeV/c, while in the beam fragmentation region it was around 10-15\% at $p_{T}>1$ GeV/c in reaction (1). In  contrary the analyzing power for reaction (2) is  almost zero at the same regions. These are the puzzling results of the (SSA) measurements in the inclusive charge exchange reaction (1) at the incident momentum of 40 GeV/c (PROZA). There is no any independent experimental confirmation of these results.

\section{First stage of the SPASCHARM polarization program}

In composing the new scientific program we are guided by the recent
theoretical and experimental developments in polarization physics. It's
obvious also that this program is also strongly influenced by our own
experiences, by resources and competitions with other collaborations over the
world. Therefore we attempt of using efficiently our proton synchrotron U70,
existing experimental  equipments and fit to the environmental requirements.
So we are going to propose the following first stage polarization program:
\begin{enumerate}
  \item
  Asymmetry measurements in charge exchange exclusive reactions at 34 GeV/c with emphasis to increase the statistics of the most of reactions shown in the Figures \ref{Fig.2} and \ref{Fig.3} by approximately by factor 10 and move to the larger –t region.

  \item
  Comparative studies of asymmetries induced by particles and antiparticles in binary and inclusive reactions.
  \item
  Study of spin transfer mechanism by using the unstable spin carrying particles like hyperons, vector mesons, etc. We emphasize, that only fixed target experiments, like ours, may measure spin transfer tensors for stable final particles, like  antiprotons and protons.
  \item
  Asymmetry measurements in inclusive productions of various stable hadrons containing partons of different flavors (u,d,s,c quarks).
  \item
  The systematic studies of the isospin dependence of single spin asymmetry.
  \item
  The comparative studies of asymmetries in production of particles and antiparticles in final state.
  \item
  Asymmetry studies by using the light ion beams and polarized target.
  \item
  Check more accurately the puzzle caused by differences of the single spin asymmetries induced by pion and proton beams in the central and fragmentation regions at 34 GeV/c.
  \item
  The new upcoming polarized proton beam will lead to the measurements of the inclusive single spin asymmetries with unprecedented precisions.
  \item
   Finally with the availability of the polarized beam and polarized target the way will be opened for the intense studies of the double spin asymmetries in many reactions.
\end{enumerate}
   For the inclusive reactions the
extensive Monte Carlo simulations were made for beam particles $\pi^{±},
K^{±}, p$ and $\bar{p}$ for momentum of 34 GeV/c. For the sake of brevity we
present, as an example, only the results of calculations for the $\bar{p}$ beam.
According to the negative beam composition the fraction of the $\bar{p}$ particles  is only 0.3\% at momentum of 34 GeV/c. Therefore the absolute flux of $\bar{p}$ beam is only $9*10^{3} \bar{p}/cycle$. The yields of the secondary particles on this beam are very important for comparison to the yields of the same particles in the proton beam. For detection of the secondary resonances with the rare decay modes produced on the propandiole target the request was imposed: the energy deposit in calorimeters should be $> 2$ GeV. Knowing the $\bar{p}$ interaction cross section with target the yields of the secondary particles with higher cross section for 30 days beam  run are presented in the next Table 1.\\
Table 1. The estimated yields $N_{EV}$ of the secondary particles from the
propandiole target ($C_{3}H_{8}O_{2}$, 20 cm long) stricken by the $\bar{p}$
beam of 34 GeV/c. One month beam run was assumed ($3.6·10^8$ interactions).
B/S means the background to signal ratio.
\begin{tabular}{|c|c|c|c|c|c|c|c|}
  \hline
  \# & particle & $N_{EV}$ &*& \#& particle & $N_{EV}$ & B/S \\
  \hline
  1 &$\pi^+$ &$2.1\times10^8$ & * & 7 & n & $1.6\times10^7 $& * \\
  \hline
  2 &$\pi^-$&$2.6\times10^8$&*&8&$\bar{n}$ &$1.4\times10^8$&* \\
   \hline
  3 &$K^+$ &$1.7\times10^7 $&* &9&$\bar{\Lambda}\rightarrow\bar{p}+\pi^+$&$2.1\times10^6$ &0.1\\
   \hline
  4 &$K^-$ &$2.2\times10^7 $&*&10&$\bar{\Lambda}\rightarrow\bar{n}+\pi^0$&$1.1\times10^6$ &8.0\\
  \hline
  5 &p&$1.6\times10^7 $&*&11 &$\bar{\Delta}^{--}\rightarrow\bar{p}+\pi^-$&$4.2\times10^7 $&7.0 \\
  \hline
  6 &$\bar{p}$ & $1.8\times10^8$& * &12 &$\Xi^{-}\rightarrow\Lambda+\pi^{-}$&$1.0\times10^5 $&0.1 \\
  \hline
\end{tabular}

Two comments are in order to this Table 1. First, the number of events for
polarized protons are less than indicated in Table 1 by one order of
magnitude. Second on the level of several percents the asymmetries produced
by protons and antiprotons may be compared if the statistics are larger than
$10^5$. It mens that such comparisons may be done for sure for pions, kaons,
barions, antibarions, but doubtful for $Xi^{-}$. The estimates were done for
ideal apparatus and not taking into accounts the real backgrounds.
%\begin{center}
\section{The experimental apparatus}
\begin{wrapfigure}[19]{LR}{180mm}
%% [number of text lines to wrap]{horizontal position: LRC}{figure width}
  \centering %% do not use \begin{center} ... \end{center}
  \vspace*{-8mm} %% the vertical position may need tweaking
  %C  \includegraphics[width=50mm]{Fig4}
  \includegraphics[width=150mm]{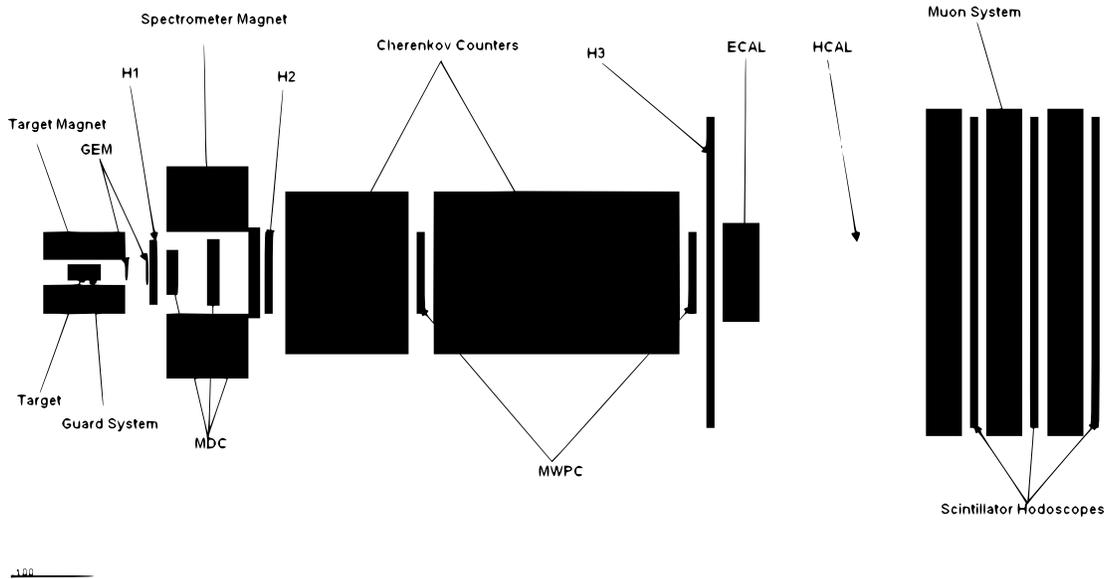}

  %  \caption{The figure caption aaaaaaa aaa, bbb bbbbbbbb.}
    \caption{Layout of the SPASCHARM experimental apparatus.}
  \label{Fig.4}
\end{wrapfigure}

%\end{center}
The experimental apparatus for the SPASCHARM program consists of the
following elements:
\begin{enumerate}
  \item
  Beam apparatus consisting of the scintillation and Cherenkov counters, scintillation hodoscopes for detecting and identifying the beam particles (not shown in Figure \ref{Fig.4}).
    \item
    The polarized proton (deuteron) target (target in Figure \ref{Fig.4}).
    \item
     The guard system surrounding the polarized target(PT).
    \item
    The polarization building-up and holding magnet (target magnet ).
    \item
     GEM1, GEM2
     \item
     The large aperture magnetic spectrometer.
      \item
      Micro drift chambers (MDC).
\item Two large aperture multichannel threshold Cherenkov counters for
identifications of the secondary particles.
   \item
   Multiwire proportional chambers (MWPC).
     \item
     Electromagnetic calorimeter(ECAL).
\item Hadron calorimeter (HCAL).
 \item
 Muon  system.
 \item
  Scintillation  hodoscopes.
\end{enumerate}
\par
The layout of the SPASCHARM detectors is presented in Figure \ref{Fig.4}. 

The main elements of the apparatus, their structure, positions and sizes are
listed in the Table 2.
\par Table 2. The parameters of the main elements of the experimental apparatus. L-distance from detector to polarized target (PT), DS-detector structure, WS-wire spacing, GS-gross size, $N_{ch}$-number of channels

\begin{tabular}{|c|c|c|c|c|c|c|}
  \hline
  detector&L, m&DS&WS&GS&$N_{ch}$ \\

  \hline
  GEM1 & 0.5 & X, Y & Strip 0.4 & 20 x 20 & 1000  \\
  \hline
  GEM2 & 0.75 & X, Y & Strip 0.4 & 30 x 30 & 1500  \\
  \hline
  1.MDC & 1.0 & X, X', Y, Y', U, V' & 6 & 65 x 54 & 1200  \\
  \hline
  2.MDC & 1.5 & X, X', Y, Y', U, V' & 6 & 111 x 81 & 1920  \\
  \hline
  3.MDC & 2.0 & X, X', Y, Y', U, V' & 6 & 150 x 111 & 2610  \\
  \hline
  4.MWPC & 3.5 & X, Y, U, V & 2 & 150 x 100 & 2500  \\
 \hline
  5.MWPC & 6.5 & X, Y, U, V & 2 & 150 x 100 & 2500  \\
 \hline 
\end{tabular}
\begin{center}
\section*{Conclusions}
\end{center}
The SPASCHARM program presents the natural extension of our previous
polarization experiments. Proposed experiment will open new and wider
perspectives due to the several reasons. First it contains the magnetic
spectrometer with the high resolution tracking detectors allowing to register
all secondary charged particles with precise angular and momentum
resolutions. Secondly it has the fast particle identification system allowing
to reconstruct the resonances with high probability. Third, the
electromagnetic and hadronic calorimeters practically allow (together with
threshold Cherenkov counters) to identify all  hadrons in final state having
sufficiently large cross sections. The large angular and momentum
acceptances will finally allow to increase by a factor 0f 10 the statistics than in previous
PROZA experiments. Using the forward detectors with the guard counters around
the polarized target one can select the binary charge exchange reactions with
one order better statistics and also detect new reactions. The detections of
hyperons and vector mesons permit to study not only polarization but also the
spin transfer mechanism in strong interaction. It is assumed that the full
apparatus for the first stage of the SPASCHARM polarization program will be
ready to 2013 beam run.
The distinct feature of our program will be the comparative studies of the polarization phenomena induced by the particles and antiparticles.\\
The work was supported by State Atomic Energy Corporation "Rosatom"
with partial support by State Agency for Science
and Innovation grant N 02.740.11.0243 and RFBR
grants 08-02-90455 and 09-02-00198.


\begin{thebibliography}{99} %% for less than 10 references use just {9}

\bibitem{gaidot1}
A.~Gaidot et al., Phys. Let. \textbf{57B} (1975)~389.
\bibitem{gaidot2}
A.~Gaidot et al., Phys. Let. \textbf{61B} (1976)~103.
\bibitem{pier1}
J.~Pierrard et al., Phys. Let. \textbf{57B} (1975)~393.
\bibitem{pier2}
J.~Pierrard et al., Phys. Let. \textbf{61B} (1976)~107.
\bibitem{mochal} V.V. Mochalov, Proc. of the 18th. Int. Spin Physics Symposium, SPIN2000, Charlottesville, Virginia, 6-11 October 2008, AIP Conf. Proc. \textbf{V.1149} (2008)pp.637-644
\bibitem{nurush}
S.B. Nurushev, Fermilab polarization experiment E581/E704: polarization
effects in pp and $\bar{p}p$ interactions at $\sqrt{s}=19.4$GeV/c.
Proceedings of the XVI-th International Seminar on High Energy Physics
Problem, June 15-20, 2002, Dubna, p.147.
\bibitem{nogach}, L.V. Nogach, Recent experimental data on spin physics. In these Proceedings.
\bibitem{xu} Qinghua Xu, Longitudinal spin transfer of Lambda and antiLambda in pp collisions at STAR. In these Proceedings.
\bibitem{logun} A. A.Logunov et al., Dokl. Akad. Nauk SSSR \textbf{142} (1962) ~317.
\bibitem{nambu} Y.Nambu, in Proceedings of International Conference on High Energy Physics, Geneva (1962), p. ~153.
\bibitem{borghini}  M. Borghini et al., Phys. Lett. \textbf{31B}  (1970) 405.
 \bibitem{bil} S. M. Bilenky et al.,, Zh.
Eksp. Teor. Fiz. \textbf{46} (1964)~1098.
\bibitem{enko} L.L. Enkovsky, B.V. Struminsky, On Polarization in the
Recharge Reaction $\pi^{-}+p\rightarrow\pi^{0}+n$. Preprint of the Institute
of Theoretical Physics, ÈÒÔ-82-160Ð, Kiev, 1982.
\bibitem{gauron}  P.  Gauron et al. Polarisation in $\pi^{-}+p\rightarrow\pi^{0}+n$ and  asymptotic theorems.Phys. Rev. Lett., 52 (1984) 1952.

\end{thebibliography}
\end{document}